\documentclass{article}

\usepackage[nonatbib,final]{nips_2016}

\pdfoutput=1

\usepackage[utf8]{inputenc} 
\usepackage[T1]{fontenc}  
\usepackage{hyperref}       
\usepackage{url}            
\usepackage{booktabs}       
\usepackage{amsfonts}       
\usepackage{nicefrac}       
\usepackage{microtype,color}      
\usepackage{bm}
\usepackage{amsmath}
\usepackage{mathtools}
\usepackage{xcolor}
\usepackage{algorithm}
\usepackage{graphicx}
\usepackage{amssymb}
\usepackage{subcaption}
\usepackage[compatible]{algpseudocode}
\usepackage[flushleft]{threeparttable}
\usepackage{cite}
\title{``Short-Dot'': Computing Large Linear Transforms Distributedly Using Coded Short Dot Products}
\author{
  Sanghamitra Dutta \\
  Carnegie Mellon University\\
  \texttt{sanghamd@andrew.cmu.edu} 
   \And
   Viveck Cadambe \\
   Pennsylvania State University\\
  \texttt{viveck@engr.psu.edu} 
   \And 
   Pulkit Grover \\
  Carnegie Mellon University\\
  \texttt{pgrover@andrew.cmu.edu} }

\begin{document}

\maketitle

\begin{abstract}
Faced with saturation of Moore's law and increasing size and dimension of data, system designers have increasingly resorted to parallel and distributed computing to reduce computation time of machine-learning algorithms. However, distributed computing is often bottle necked by a small fraction of slow processors called ``stragglers'' that reduce the speed of computation because the fusion node has to wait for all processors to complete their processing. To combat the effect of stragglers, recent literature proposes introducing redundancy in computations across processors, e.g.,~using repetition-based strategies or erasure codes. The fusion node can exploit this redundancy by completing the computation using outputs from only a subset of the processors, ignoring the stragglers. In this paper, we propose a novel technique -- that we call ``Short-Dot'' -- to introduce redundant computations in a coding theory inspired fashion, for computing linear transforms of long vectors. Instead of computing long dot products as required in the original linear transform, we construct a larger number of redundant and short dot products that can be computed faster and more efficiently at individual processors. In reference to comparable schemes that introduce redundancy to tackle stragglers,  Short-Dot reduces the cost of computation, storage and communication since shorter portions are stored and computed at each processor, and also shorter portions of the input is communicated to each processor. Further, only a subset of these short dot products are required at the fusion node to finish the computation successfully, thus enabling us to ignore stragglers. We demonstrate through probabilistic analysis as well as experiments on computing clusters that Short-Dot offers significant speed-up compared to existing techniques. We also derive trade-offs between the length of the dot-products and the resilience to stragglers (number of processors to wait for), for any such strategy and compare it to that achieved by our strategy. 
\end{abstract}

\section{Introduction}
\label{sec:introduction}
This work proposes a coding-theory inspired computation technique for speeding up computing linear transforms of high-dimensional data by distributing it across multiple processing units that compute shorter dot products. Our main focus is on addressing the ``straggler effect,'' \textit{i.e.}, the problem of delays caused by a few slow processors that bottleneck the entire computation. To address this problem, we provide techniques (building on~\cite{kananspeeding}~\cite{gauristraggler}~\cite{gauriefficient}~\cite{gauri2014delay}~\cite{huang2012codes}) that introduce redundancy in the computation by designing a novel error-correction mechanism that allows the size of individual dot products computed at each processor to be shorter than the length of the input. Shorter dot products offer advantages in computation, storage and communication in distributed linear transforms.

The problem of computing linear transforms of high-dimensional vectors is ``the" critical step \cite{dally2015} in several machine learning and signal processing applications. Dimensionality reduction techniques such as Principal Component Analysis (PCA), Linear Discriminant Analysis (LDA), taking random projections, require the computation of short and fat linear transforms on high-dimensional data. Linear transforms are the building blocks of solutions to various machine learning problems, e.g., regression and classification etc., and are also used in acquiring and pre-processing the data through Fourier transforms, wavelet transforms, filtering, etc. Fast and reliable computation of linear transforms are thus a necessity for low-latency inference \cite{dally2015}. Due to saturation of Moore's law, increasing speed of computing in a single processor is becoming difficult, forcing practitioners to adopt parallel processing to speed up computing for ever increasing data dimensions and sizes.

Classical approaches of computing linear transforms across parallel processors, e.g., Block-Striped Decomposition~\cite{kumar1994introduction}, Fox's method \cite{fox1987matrix,kumar1994introduction}, and Cannon's method \cite{kumar1994introduction}, rely on dividing the computational task equally among all available processors\footnote{Strassen's algorithm \cite{strassen1969gaussian} and its generalizations offer a recursive approach to faster matrix multiplications over multiple processors, but they are often not preferred because of their high communication cost \cite{ballard2014communication}.} without any redundant computation. The fusion node collects the outputs from each processors to complete the computation and thus has to wait for all the processors to finish. In almost all distributed systems, a few slow or faulty processors -- called ``stragglers''\cite{straggler_tail} -- are observed to delay the entire computation. This unpredictable latency in distributed systems is attributed to factors such as network latency, shared resources, maintenance activities, and power limitations. In order to combat with stragglers, cloud computing frameworks like Hadoop~\cite{hadoop} employ various straggler detection techniques and usually reset the task allotted to stragglers. Forward error-correction techniques offer an alternative approach to deal with this ``straggler effect'' by introducing  redundancy in the computational tasks across different processors. The fusion node now requires outputs from only a subset of all the processors to successfully finish. In this context, the use of preliminary erasure codes dates back to the ideas
of algorithmic fault tolerance \cite{ABFT1984}~\cite{faultbook}. Recently optimized Repetition and  Maximum Distance Separable (MDS)~\cite{ryan2009channel} codes have been explored  \cite{gauristraggler}~ \cite{gauriefficient}~~\cite{kananspeeding}~\cite{mohammad2016}  to speed up computations. 

We consider the problem of computing $\bm{A}\bm{x}$ where $\bm{A}_{(M \times N)}$ is a given matrix and $\bm{x}_{( N \times 1)}$ is a vector that is input to the computation $(M\ll N)$. In contrast with~\cite{kananspeeding}, which also uses codes to compute linear transforms in parallel, we allow the size of individual dot products computed at each processor to be smaller than $N$, the length of the input. 

\textbf{Why might one be interested in computing short dot products while performing an overall large linear transform?} \\
One reason is straightforward: the computation time depends on the length of the dot-products computed. Processors are also inherently memory limited, which limits the size of dot products that can be computed. In some distributed and cloud computing systems, the computation time is dominated by the time taken to communicate $\bm{x}$ to the processors. In systems where multi-casting $\bm{x}$ is not possible or is inefficient, it may be faster to communicate a subset of the co-ordinates of $\bm{x}$ to each processor.  In such systems, we anticipate that communicating shorter vectors, each formed by these subsets of coordinates of $\bm{x}$, is likely to result in substantial speedups over schemes that require the entire $\bm{x}$ vector (in particular when multi-casting is difficult).\footnote{Another interesting example comes from recent work on designing processing units that exclusively compute dot-products using analog components~\cite{analog_dot,ericpop}. These devices are prone to errors and increased delays in convergence when designed for larger dot products.} In Sections~\ref{sec:analysis} and~\ref{sec:experiments}, we show both theoretically (under model assumptions inspired from~\cite{kananspeeding} that admit simplified expected time analysis while being a crude approximation of experimental observations) and experimentally that the speed-up using Short-Dot can be increased beyond that obtained using the strategy proposed in ~\cite{kananspeeding}, in straggler-prone environments.


To summarize, our main contributions are:

\begin{enumerate}
\item To compute $\bm{A} \bm{x}$ for a given matrix $\bm{A}_{(M \times N)}$, we instead compute $\bm{F} \bm{x}$ where we construct $\bm{F}_{(P \times N)}$ (total no. of processors $P$ > Required no. of dot-products $M$) such that each $N$-length row of $\bm{F}$ has at most $ N(P-K+M)/P$ non-zero elements. Because the locations of zeros in a row of $\bm{F}$ are known by design, this reduces the complexity of computing dot-products of rows of $\bm{F}$ with $\bm{x}$. Here $K$ parameterizes the resilience to stragglers: any $K$ of the $P$ dot products of rows of $\bm{F}$ with $\bm{x}$ are sufficient to recover $\bm{A}\bm{x}$, \textit{i.e.}, any $K$ rows of $\bm{F}$ can be linearly combined to generate the rows of $\bm{A}$.\\
\item We provide fundamental limits on the trade-off between the length of the dot-products and the straggler resilience (number of processors to wait for) for \textit{any} such strategy in Section 3. This suggests a lower bound on the length of task allotted per processor. Our limits show that Short-Dot is near-optimal.\\
\item Assuming exponential tails of service-times at each server (used in~\cite{kananspeeding}), we derive the expected computation time required by our strategy and compare it to uncoded parallel processing, repetition strategy and MDS coding~\cite{ryan2009channel} (see Fig.~\ref{multiavp}) based linear computation. We also explicitly show a regime ($M=\frac{P}{\log P}$) where Short-Dot outperforms all its competing strategies in expected computation time, by a factor of $\frac{\log(P)}{\log( \log P)}$, that diverges to infinity for large $P$. In general, Short-Dot is found to be universally faster than all its competing strategies over the entire range of $M \leq P$. When $M$ is linear in $P$, Short-Dot offers speed-up by a factor of $\Omega(\log(P))$ over uncoded, parallel processing and repetition. When $M$ is sub-linear in $P$, Short-Dot out-performs repetition or MDS coding based linear computations by a factor of $\Omega \left(\frac{P}{M\log(P/M)}\right)$ .\\
\item We also provide experimental results showing that Short-Dot is faster than existing strategies.
\end{enumerate}

%


In a concurrent work, in \cite{gradientcoding}, Tandon et al. consider a coded computation problem similar to ours for the special case where $M$, the number of $N$-length dot-products to be computed, is $1$ and the given matrix $\bm{A}_{M \times N}$ (in this case just a single row vector) is $[1, 1, \dots , 1]_{1 \times N}$. We note that for $M=1$, the gain of using coded strategies over replication-based strategies is bounded even as $N$ and $P \to\infty$ for $s=\Theta(N/P) $ . Our paper differs from \cite{gradientcoding} in that we consider the more general case $M\geq 1$, and observe that the gains over replication can be unbounded with this scaling in the regime $s=\Theta(MN/P) $. For $M>1$, the number of operations per processor using our strategy is lower than an application of \cite{gradientcoding} for the same worst-case straggler resilience. To see this, note that a straightforward extension of the strategy proposed in \cite{gradientcoding} that encodes each row of $\bm{A}$ separately for $M (> 1)$ rows would require $M$ dot-products of length $\frac{N(P-K+1)}{P}$ at each processor while using a ``joint'' encoding across rows, Short-Dot only requires a single dot-product of length $\frac{N(P-K+M)}{P}$ \big(note that $ \frac{N(P-K+M)}{P} < M \times \frac{N(P-K+1)}{P}$\big)  at each processor, while still requiring the same number of processors (any $K$ out of $P$) to finish. Further, we also provide a tighter converse for $M > 1$ that proves that Short-Dot is near-optimal. It is worth noting that \cite{gradientcoding} also introduces the notion of partial stragglers, which is outside the scope of our paper.

For the rest of the paper, we define the sparsity of a vector $\bm{u} \in \mathbb{R}^N$ as the number of nonzero elements in the vector, \textit{i.e.}, $ \|\bm{u}\|_0 = \sum_{j=1}^N \mathcal{I}(u_j \neq 0) $. We also assume $N$ is quite large compared to $P$, so that it is reasonable to assume that $P$ divides $N$ ($P \ll N$). 
\subsection{Comparison with existing strategies:}

\begin{figure}[t]
\centering
\begin{minipage}[b]{0.45\linewidth}
\centering
\fbox{\includegraphics[width=6.2cm, height=3.2cm]{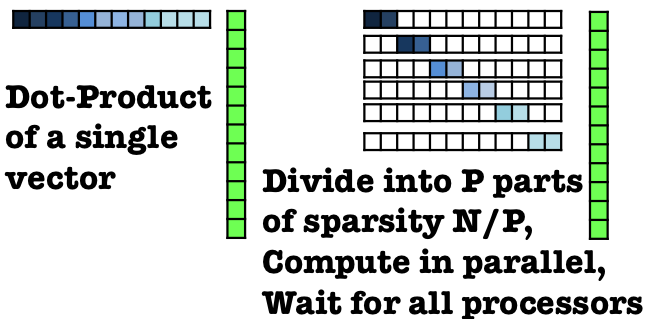}}
\end{minipage}
\begin{minipage}[b]{0.45\linewidth}
\centering
\fbox{\includegraphics[width=6.2cm, height=3.2cm]{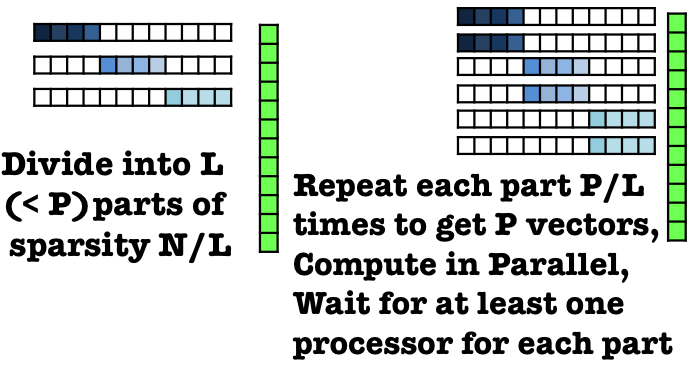}}
\end{minipage}
\caption{ A dot-product of length $N=12$ is being computed parallely using $P=6$ processors. (Left) Uncoded Parallel Processing - Divide into $P$ parts, (Right) Repetition with block partitioning.}\label{single_dot}
\end{figure}

Consider the problem of computing a single dot product of an input vector $\bm{x} \in \mathbb{R}^N$ with a pre-specified vector $\bm{a} \in \mathbb{R}^N$. By an ``uncoded'' parallel processing strategy (which includes Block Striped Decomposition \cite{kumar1994introduction}), we mean a strategy that does not use redundancy to overcome delays caused by stragglers. One uncoded strategy is to partition the dot product into $P$ smaller dot products, where $P$ is the number of available processors. E.g. $\bm{a}$ can be divided into $P$ parts -- constructing $P$ short vectors of sparsity $N/P$ -- with each vector stored in a different processor (as shown in Fig. \ref{single_dot} left). Only the nonzero values of the vector need to be stored since the locations of the nonzero values is known apriori at every node. One might expect the computation time for each processor to reduce by a factor of $P$. However, now the fusion node has to wait for all the $P$ processors to finish their computation, and the stragglers can now delay the entire computation. Can we construct $P$ vectors such that dot products of a subset of them with $\bm{x}$ are sufficient to compute $\langle \bm{a},\bm{x} \rangle $? A simple coded strategy is Repetition with block partitioning \textit{i.e.}, constructing $L$ vectors of sparsity $N/L$ by partitioning the vector of length $N$ into $L$ parts $(L < P)$, and repeating the $L$ vectors $P/L$ times so as to obtain $P$ vectors of sparsity $N/L$ as shown in Fig.~\ref{single_dot} (right). For each of the $L$ parts of the vector, the fusion node only needs the output of one processor among all its repetitions. Instead of a single dot-product, if one requires the dot-product of $\bm{x}$ with $M$ vectors $\{\bm{a}_1,\ldots,\bm{a}_M\}$, one can simply repeat the aforementioned strategy $M$ times. 

\begin{figure}[ht]

\begin{minipage}[b]{0.5\linewidth}
\centering
\fbox{\includegraphics[width=6cm,height=4.5cm]{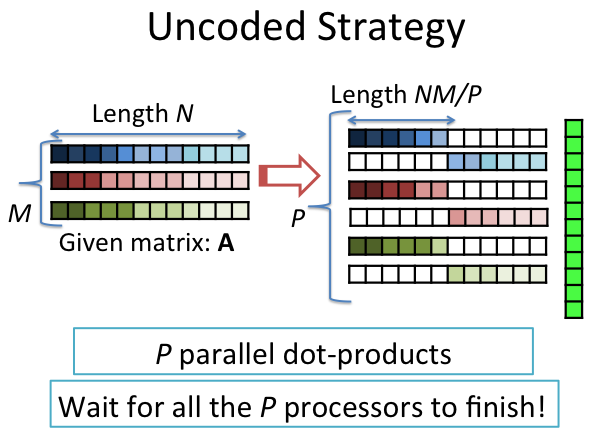}}
\subcaption{Uncoded Parallel Processing}\label{fig:1b}
\end{minipage}
\begin{minipage}[b]{0.47\linewidth}
\fbox{\includegraphics[width=6cm,height=4.5cm]{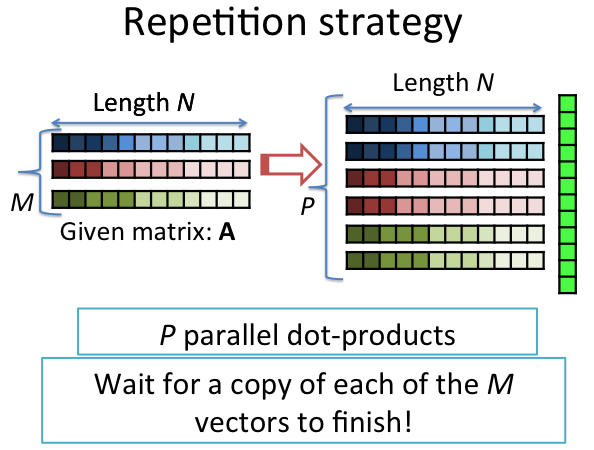}}
\subcaption{Repetition Strategy}\label{fig:1c} 
\end{minipage}
\begin{minipage}[b]{0.5\linewidth}
\centering
\fbox{\includegraphics[width=6cm,height=4.5cm]{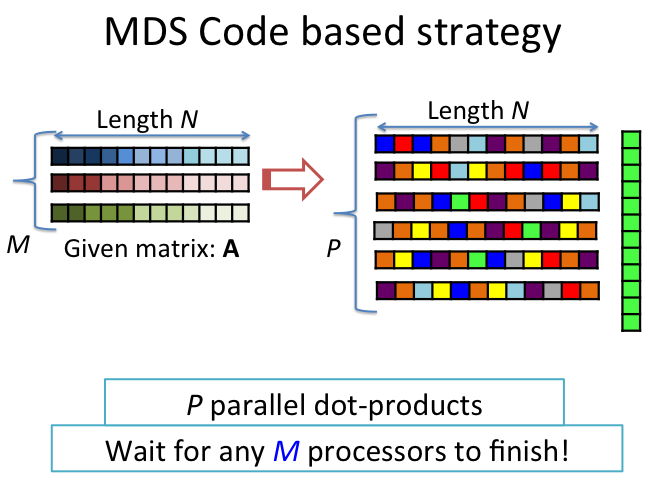}}
\subcaption{Using MDS codes}\label{fig:1d}
\end{minipage}
\begin{minipage}[b]{0.48\linewidth}
\centering
\fbox{\includegraphics[width=6.1cm, height=4.5cm]{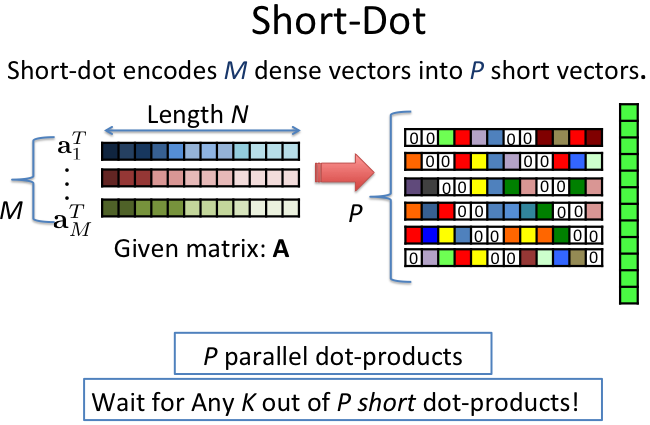}}
\subcaption{Using Short-Dot}\label{fig:1e}
\end{minipage}
\caption{Different strategies of parallel processing: Here $M=3$ dot-products of length $N=12$ are being computed using $P=6$ processors.}\label{multiavp}
\end{figure}

For multiple dot-products, an alternative repetition-based strategy is to compute $M$ dot products $P/M$ times in parallel at different processors. Now we only have to wait for at least one processor corresponding to each of the $M$ vectors to finish (see Fig.~\ref{fig:1c}). Improving upon repetition, it is shown in \cite{kananspeeding} that an $(P,M)$-MDS code allows constructing $P$ coded vectors such that any $M$ of $P$ dot-products can be used to reconstruct all the $M$ original vectors (see Fig.~\ref{fig:1d}). This strategy is shown, both experimentally and theoretically, to perform better than repetition and uncoded strategies.

\textit{Can we go beyond MDS codes?} MDS codes-based strategies require $N$-length dot-products to be computed on each processor. Short-Dot instead constructs $P$  vectors of sparsity $s$ (less than $N$), such that the dot product of $\bm{x}$ with any $K \ ( \geq M)$ out of these $P$ short vectors is sufficient to compute the dot-product of $\bm{x}$ with all the $M$ given vectors (see Fig.~\ref{fig:1e}). Compared to MDS Codes, Short-Dot is more flexible as it waits for some more processors (since $K \geq M$), but each processor computes a shorter dot product. Short-Dot also effectively reduces the communication cost since only a shorter portion of the input vector is to be communicated to each processor. We also propose Short-MDS, an extension of the MDS codes-based strategy in \cite{kananspeeding} to create short dot-products of length $s$, through block partitioning, and compare it with Short-Dot. In regimes where $\frac{N}{s}$ is an integer, Short-MDS may be viewed as a special case of Short-Dot. But when $\frac{N}{s}$ is not an integer, Short-MDS has to wait for more processors in worst case than Short-Dot for the same sparsity $s$, as discussed in Remark 2 in Section~\ref{sec:shortdot}.



\section{Our coded parallelization strategy: Short-Dot}\label{sec:shortdot}

In this section, we provide our strategy of computing the linear transform $\bm{A}\bm{x}$ where $\bm{x} \in \mathbb{R}^N$ is the input vector and $\bm{A}_{(M \times N)}=[\bm{a}_1,\bm{a}_2,\ldots,\bm{a}_M]^T$ is a given matrix. Short-Dot constructs a $P \times N$ matrix $\bm{F}=[\bm{f}_1,\bm{f}_2,\ldots,\bm{f}_P]^T$  such that $M$ predetermined linear combinations of \textit{any} $K$ rows of $\bm{F}$ are sufficient to generate each of $\{ \bm{a}_1^T, \ldots, \bm{a}_M^T \}$, and any row of $\bm{F}$ has sparsity at most $s = \frac{N}{P}(P-K+M)$. Each sparse row of $\bm{F}$ (say $\bm{f}_i^T$) is sent to the $i$-th processor ($i=1,\ldots,P$) and dot-products of $\bm{x}$ with all sparse rows are computed in parallel. Let $S_i$ denote the support (set of non-zero indices) of $\bm{f}_i$. Thus, for any unknown vector $\bm{x}$, short dot products of length $|S_i| \leq s = \frac{N}{P}(P-K+M)$ are computed on each processor. Since the linear combination of any $K$ rows of $\bm{F}$ can generate the rows of $\bm{A}$, \textit{i.e.}, $\{ \bm{a}_1^T, \bm{a}_2^T, \ldots, \bm{a}_M^T \}$, the dot-product from the earliest $K$ out of $P$ processors can be linearly combined to obtain the linear transform $\bm{A}\bm{x}$.
Before formally stating our algorithm, we first provide an insight into why such a matrix $\bm{F}$ exists in the following theorem, and develop an intuition on the construction strategy.
\begin{figure}[ht]
\centering
\fbox{\includegraphics[height=3.4cm]{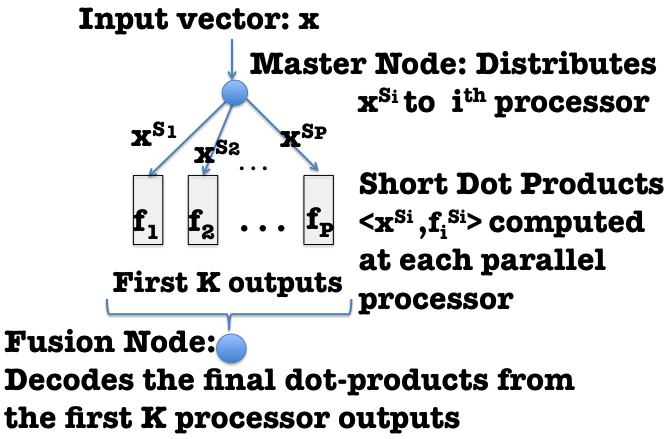}}
\caption{Short-Dot: Distributes short dot-products over $P$ parallel processors, such that outputs from any $K$ out of $P$ processors are sufficient to compute successfully.}
\end{figure}


\newtheorem{theorem}{Theorem}
\newtheorem{lemma}{Lemma}

\begin{theorem}
Given row vectors $\{ \bm{a}_1^T, \bm{a}_2^T, \ldots, \bm{a}_M^T \}$, there exists a $P \times N$ matrix $\bm{F}$ such that a linear combination of \textit{any} $K (>M)$ rows of the matrix is sufficient to generate the row vectors and each row of $\bm{F}$ has sparsity at most $s = \frac{N}{P}(P-K+M) $, provided $P$ divides $N$.
\end{theorem}
\textit{Proof:}
We may append $(K-M)$ rows to $\bm{A}=[\bm{a}_1,\bm{a}_2,\ldots,\bm{a}_M]^T $, to form a $K \times N$ matrix $\bm{\tilde{A}}=[\bm{a}_1,\bm{a}_2,\ldots,\bm{a}_M,\bm{z}_{1},\ldots,\bm{z}_{K-M}]^T$. The precise choice of these additional vectors will be made explicit later. Next, we choose $\bm{B}$, a $P \times K$ matrix such that \textit{any square sub-matrix of $\bm{B}$ is invertible}\footnote{This condition is relaxed in Remark 1.} The following lemma shows that \textit{any} $K$ rows of the matrix $\bm{B\tilde{A}}$ are sufficient to generate any row of $\bm{\tilde{A}}$, including $\{\bm{a}_1^T,\bm{a}_2^T,\ldots,\bm{a}_M^T\}$:
\begin{lemma} Let $\bm{F}=\bm{B\tilde{A}}$ where $\bm{\tilde{A}}$ is a $K \times N$ matrix and $\bm{B}$ is any $(P\times K)$ matrix such that every square sub-matrix is invertible. Then, any $K$ rows of $\bm{F}$ can be linearly combined to generate any row of $\bm{\tilde{A}}$.
\end{lemma}
\textit{Proof:} Choose an arbitrary index set $\chi \subset \{1,2,\ldots, P\} $ such that $|\chi|=K$.  Let $\bm{F}^{\chi}$ be the sub-matrix formed by chosen $K$ rows of $\bm{F}$ indexed by $\chi$. Then, $\bm{F}^{\chi}=\bm{B}^{\chi}\bm{\tilde{A}}$. Now, $\bm{B}^{\chi}$ is a $K \times K$ sub-matrix of $\bm{B}$, and is thus invertible. Thus, $ \bm{\tilde{A}}=(\bm{B}^{\chi})^{-1}\bm{F}^{\chi} $. The $i$-th row of $\bm{\tilde{A}}$ is $[i$-th Row of  $(\bm{B}^{\chi})^{-1}]\bm{F}^{\chi}$ for $i=1,2,\ldots,K$. Thus, each row of $\bm{\tilde{A}}$ is generated by the chosen $K$ rows of $\bm{F}$.  \hfill $\blacksquare$

In the next lemma, we show how the row sparsity of $\bm{F}$ can be constrained to be at most $\frac{N}{P}(P-K+M)$ by appropriately choosing the appended vectors $\bm{z}_{1},\ldots,\bm{z}_{K-M}$.
\begin{lemma} Given an $M \times N$ matrix $\bm{A}=[\bm{a}_1,\ldots,\bm{a}_M]^T$, let $\bm{\tilde{A}}=[\bm{a}_1,\ldots,\bm{a}_M,\bm{z}_{1},\ldots,\bm{z}_{K-M}]^T$ be a $K \times N$ matrix formed by appending $K-M$ row vectors to $\bm{A}$. Also let $\bm{B}$ be a $P \times K$ matrix such that every square matrix is invertible. Then there exists a choice of the appended vectors $\bm{z}_{1},\ldots,\bm{z}_{K-M}$ such that each row of $\bm{F} = \bm{B}\bm{\tilde{A}}$ has sparsity at most $s =\frac{N}{P}(P-K+M)$. 
\end{lemma}

\begin{figure}[H]
\centering
\fbox{\includegraphics[width=6.2cm, height=3.1cm]{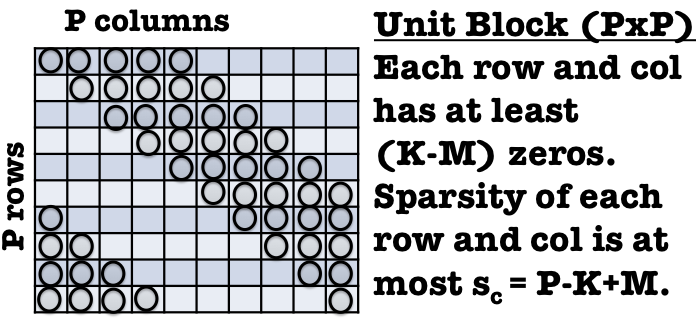}}
\hspace{0.5cm}
\fbox{\includegraphics[width=5.6cm, height=3.1cm]{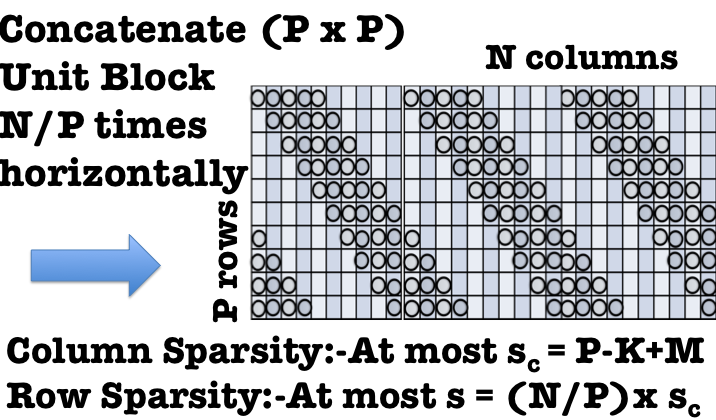}}
\caption{Sparsity pattern of $\bm{F}$: (Left) Unit Block $(P \times P)$; (Right) Unit Block concatenated $N/P$ times to form $N \times P$ matrix $\bm{F}$ with row sparsity at most $s$.}\label{sparsity_fig}
\end{figure}

\textit{Proof:} We select a sparsity pattern that we want to enforce on $\bm{F}$ and then show that there exists a choice of the appended vectors $\bm{z}_{1},\ldots,\bm{z}_{K-M}$ such that the pattern can be enforced.\\
\textbf{Sparsity Pattern enforced on $\bm{F}$:} This is illustrated in Fig.~\ref{sparsity_fig}. First, we construct a $P \times P$ ``unit block'' with a cyclic structure of nonzero entries, where $(K-M)$ zeros in each row and column are arranged as shown in Fig. \ref{sparsity_fig}. Each row and column have at most $s_c= P-K+M$ non-zero entries. This unit block is replicated horizontally $N/P$ times to form an $P \times N$ matrix with at most $s_c$ non-zero entries in each column, and and at most $s=Ns_r/P$ non-zero entries in each row. We now show how choice of $\bm{z}_{1},\ldots,\bm{z}_{K-M}$ can enforce this pattern on $\bm{F}$.

From $\bm{F}=\bm{B}\bm{\tilde{A}} $, the $j$-th column of $\bm{F}$ can be written as, $\bm{F}_j=\bm{B}\bm{\tilde{A}}_j $. Each column of $\bm{F}$ has at least $K-M$ zeros at locations indexed by $U \subset \{1,2,\ldots, P\}$. Let $\bm{B}^U$ denote a $( (K-M) \times K )$ sub-matrix of $\bm{B}$ consisting of the rows of $\bm{B}$ indexed by $U$. Thus, $   \bm{B}^U \bm{\tilde{A}}_j= [\bm{0}]_{(K-M) \times 1} $.

Divide $\bm{\tilde{A}}_j$ into two portions of lengths $M$ and $K-M$ as follows: 

$\bm{\tilde{A}}_j =[\bm{A}_j^T \ \ | \ \bm{z}^T]^T \ =[ a_1(j) \ a_2(j) \ldots a_M(j) \  \ \ z_1(j)\ \ldots \ z_{K-M}(j)]^T $ 

 Here $\bm{A}_j=
[ a_1(j) \ a_2(j) \ldots a_M(j)]^T $ is actually the $j$-th column of given matrix $\bm{A}$ and $\bm{z} =[ z_{1}(j),\ \ldots \ z_{K-M}(j)]^T$ depends on the choice of the appended vectors. Thus,
\begin{align}
& \nonumber \bm{B}^U_{cols \ 1:M}\bm{A}_j  + \bm{B}^U_{cols \ M+1:K}\ \bm{z}\  = [\bm{0}]_{K-M \times 1}  \\
 \Rightarrow  & \bm{B}^U_{cols \ M+1:K}\ \bm{z}\  = - \bm{B}^U_{cols \ 1:M}[\bm{A}_j] \\  \Rightarrow & [\ \bm{z}\ ]=  -  (\bm{B}^U_{cols \ M+1:K})^{-1} \ \bm{B}^U_{cols \ 1:M}[\bm{A}_j]
\end{align}
where the last step uses the fact that $[\bm{B}^U_{cols \ M+1:K}]$ is invertible because it is a $(K-M) \times (K-M)$ square sub-matrix of $\bm{B}$. This explicitly provides the vector $\bm{z}$ which completes the $j$-th column of $\tilde{\bm{A}}$. The other columns of $\tilde{\bm{A}}$ can be completed similarly, proving the lemma. \hfill $\blacksquare$\\

From Lemmas 1 and 2, for a given $M \times N$ matrix $\bm{A}$, there always exists a $P \times N$ matrix $\bm{F}$ such that a linear combination of \textit{any} $K$ columns of $\bm{F}$ is sufficient to generate our given vectors  and each row of $\bm{F}$ has sparsity at most $s = \frac{N}{P}(P-K+M) $. This proves the theorem.\hfill $\blacksquare$ \\

\textbf{Remark 1: Relaxed conditions on matrix $\bm{B}$} 

It has been stated in \textit{Lemmas} $1$ and $2$ that all square sub-matrices of $\bm{B}$ need to be invertible. A matrix with \textit{i.i.d.}  Gaussian entries can be shown to satisfy this property with probability $1$. In fact the condition on $\bm{B}$ in \textit{Lemmas} $1$ and $2$  can be relaxed, as evident from the proof. For matrix $\bm{B}_{P \times K}$ we only need two conditions. (1) All $K \times K$ square sub-matrices are invertible. (2) All $(K-M) \times (K-M)$ square sub-matrices in the last $K-M$ columns of $\bm{B}$ are invertible. A Vandermonde Matrix satisfies both these properties and thus can be used for encoding in Short-Dot.

With this insight in mind, we now formally state our computation strategy:

\begin{algorithm}[H]
\caption{Short-Dot}
\label{algo}
\begin{algorithmic}
\STATE \textbf{[A] Pre-Processing Step: Encode $\bm{F}$ (Performed Offline)}
\STATE \textbf{Given:}  $ \bm{A}_{M \times N}=[\bm{a}_1,\ldots,\bm{a}_M]^T=[\bm{A}_1, \bm{A}_2,\ldots, \bm{A}_N],\ \ parameter\ K , Matrix \ \bm{B}_{P \times K}$
\end{algorithmic}
\begin{algorithmic}[1]
\STATE \textbf{For} $ \ j = 1 \ to \ N \ $ \textbf{ do }
\State \hspace{0.5cm} \textbf{Set } $ \ \ \  \ U \gets (\{ (j-1), \ldots, (j+K-M-1) \} \mod P) +1 $\\
 \hfill $\rhd$ The set of $(K-M)$ indices that are 0 for the $j$-th column of $\bm{F}$
\State \hspace{0.5cm} \textbf{Set} $\ \ \ \bm{B}^U \gets \text{Rows of } \bm{B} \text{ indexed by $U$}$
\State \hspace{0.5cm} \textbf{Set} $ \ \ \ \ \  [\ \bm{z}\ ]=  -  (\bm{B}^U_{cols \ M+1:K})^{-1} \ \bm{B}^U_{cols \ 1:M}[\bm{A}_j]  $  \hfill $\rhd$ $\bm{z}_{(K-M) \times 1}$ is a row vector.
\State \hspace{0.5cm} \textbf{Set} $ \ \ \ \ \ \bm{F}_{j} = \bm{B}[\bm{A}_j^T \ | \bm{z}^T \ ]^T $ \hfill $\rhd$ $\bm{F}_{j}$ is a column vector ( $j$-th col of $\bm{F}$)
\Statex \textbf{Encoded Output:} $\bm{F}_{P \times N}=[\bm{f}_1 \bm{f}_2 \ldots \bm{f}_P]^T$ \hfill $\rhd$ Row representation of matrix $\bm{F}$
\STATE \textbf{For} $ \ i = 1 \ to \ P \ $ \textbf{ do }
\State \hspace{0.5cm} \textbf{Store}  $ \ \ \ S_i \gets Support(\bm{f}_i)$
\hfill $\rhd$ Indices of non-zero entries in the $i$-th row of $\bm{F}$
\State \hspace{0.5cm} \textbf{Send} $ \ \ \ \bm{f}_i^{S_i}$ to $i$-th processor  \hfill $\rhd$ $i$-th row of $\bm{F}$ sent to $i$-th processor
\end{algorithmic}
\begin{algorithmic}
\State \textbf{[B] Online computations}
\State \textbf{External Input :}  $  \bm{x}$
\State \textbf{Resources: } $ P$ parallel processors $(P >M)$
\State \textbf{[B1] Parallelization Strategy: Divide task among parallel processors:}
\end{algorithmic}
\begin{algorithmic}[1]
\STATE \textbf{For} $ \ i = 1 \ to \ P \ $ \textbf{ do }
\State \hspace{0.5cm} {Send $\bm{x}^{S_i}$ to the $i$-th processor}
\State \hspace{0.5cm} {Compute at $i$-th processor:} $\langle \bm{f}_i^{S_i} , \bm{x}^{S_i} \rangle$
\hfill $\rhd$ $\bm{u}^S$ denotes only the rows of vector $\bm{u}$ indexed by $S$
\end{algorithmic}
\begin{algorithmic}
\State \textbf{Output:} $\langle \bm{f}_i^{S_i} , \bm{x}^{S_i} \rangle$ from $K$ earliest processors 
\algstore{algo2}
\end{algorithmic}
\end{algorithm}

\begin{algorithm}
\begin{algorithmic}
\algrestore{algo2}
\State \textbf{[B2] Fusion Node: Decode the dot-products from the processor outputs:}
\end{algorithmic}
\begin{algorithmic}[1]
\State  \textbf{Set }  $\ \ \  \ V \gets \text{Indices of the } $K$ \text{ processors that finished first}$
\State  \textbf{Set} $\ \ \ \ \bm{B}^V \gets \text{Rows of } \bm{B} \text{ indexed by $V$}$
\State  \textbf{Set} $\ \ \ \ \bm{v}_{K \times 1} \gets [\langle \bm{f}_i^{S_i} , \bm{x}^{S_i} \rangle , \ \forall \ i \ \in V]$ \hfill $\rhd$ Col Vector of outputs  from first $ K $ processors
\State \textbf{Set  } $\bm{Ax} = [\langle \bm{a}_1,\bm{x} \rangle, \ldots,  \langle\bm{a}_M,\bm{x}\rangle]^T \gets [(\bm{B}^V)^{-1}]^{rows \ 1:M} \bm{v}$
\State \textbf{Output:} $\langle \bm{x} ,\bm{a}_1 \rangle, \ldots,  \langle\bm{x}, \bm{a}_M\rangle $
\end{algorithmic}
\end{algorithm}
\begin{table}[ht]
\centering
  \caption{Trade-off between the length of the dot-products and parameter $K$ for different strategies}
  \label{results_1}
  \hspace{1mm}
\begin{tabular}{lll}
\midrule
Strategy& Length& Parameter $K$\\
\midrule
Repetition & $N$ &  $P- \left \lfloor{\frac{P}{M}}\right \rfloor +1 $ \\
\midrule
MDS & $N$ &  $M$ \\
\hline
Short-Dot & $s$ & $P-\left \lfloor{\frac{Ps}{N}}\right \rfloor+M$   \\
\hline
\end{tabular}
\quad
\begin{tabular}{p{22mm}p{9mm}p{28mm}}
\midrule
Strategy & Length & Parameter $K$  \\
  \midrule
Repetition with block partition & $s$ &  $ P-\left \lfloor{\frac{P}{M\left \lceil{N/s}\right \rceil}}\right \rfloor+1$ \\
\hline
Short-MDS & $s$ & $ P-\left \lfloor{\frac{P}{\left \lceil{N/s}\right \rceil}}\right \rfloor+M$   \\
\hline
\end{tabular}
\end{table}

\textbf{Remark 2: Short-MDS - a special case of Short-Dot} 

An extension of the MDS codes-based strategy proposed in \cite{kananspeeding}, that we call Short-MDS can be designed to achieve row-sparsity $s$. First block-partition the matrix of $N$ columns, into $\left \lceil{N/s}\right \rceil$ sub-matrices of size $M \times s$, and also divide the total processors $P$ equally into $\left \lceil{N/s}\right \rceil$ parts. Now, each sub-matrix can be encoded using a $(\frac{P}{\left \lceil{N/s}\right \rceil},M)$ MDS code. In the worst case, including all integer effects, this strategy requires $K=P-\left \lfloor{\frac{P}{\left \lceil{N/s}\right \rceil}}\right \rfloor+M$ processors to finish. In comparison, Short-Dot requires $K=P- \left \lfloor{\frac{Ps}{N}}\right \rfloor+M$ processors to finish. In the regime where, $s$ exactly divides $N$, Short-MDS can be viewed as a special case of Short-Dot, as both the expressions match. However, in the regime where $s$ does not exactly divide $N$, Short-MDS requires more processors to finish in the worst case than Short-Dot. Short-Dot is a generalized framework that can achieve a wider variety of pre-specified sparsity patterns as required by the application. In Table~\ref{results_1}, we compare the lengths of the dot-products and straggler resilience $K$, \textit{i.e.}, the number of processors to wait for in worst case, for different strategies.

\section{Limits on trade-off between the length of dot-products and parameter \textit{K}} 
\label{sec:fundamental}

In this section, we derive fundamental trade-offs between the length of the dot-products computed at each individual processor and the number of processors to wait for, \textit{i.e.}, $K$, which parametrizes the resilience to stragglers. First we derive an information-theoretic limit in Theorem \ref{thm:fundamental1} that holds for any matrix $\bm{A}$, such that each column has at least one non-zero entry\footnote{Note that choice of such a class of matrix $\bm{A}$ is reasonable, since if say the $j$-th column of $\bm{A}$ consists entirely of zeros, then the $j$-th column and its corresponding entry in unknown vector $\bm{x}$ can simply be omitted from the problem.}. In Theorem \ref{thm:fundamental2}, we show how this bound can be tightened further, so that in the limit of large number of columns of matrix $\bm{A}$, Short-Dot is near-optimal. 
\begin{theorem}
\label{thm:fundamental1}
Let $\bm{A}_{M \times N}$ be any matrix such that each column has at least one non-zero element. For any matrix $\bm{F}_{P \times N}$ satisfying the property that the span of its \textit{any} $K$ rows contains the span of the $M$ rows of  $\bm{A}_{M \times N}$,  the average sparsity $\bar{s}$ over the rows of $\bm{F}_{P \times N}$ must satisfy 
$\bar{s} \geq \frac{N}{P}\big(P-K + 1\big) $.
\end{theorem}

\textit{Proof:} We claim that $K$ is strictly greater than the maximum number of zeros that can occur in any column of the matrix $\bm{F}$. If not, suppose the $j$-th column of $\bm{F}$ has more than $K$ zeros. Then there exists a choice of $K$ rows of $\bm{F}$ such that any linear combination of these rows will always be $0$ at the $j$-th column index. However, since the $j$-th column of $\bm{A}$ has at least one non-zero entry, say at row $i$, it is not possible to generate the $i$-th row of $\bm{A}$ by linearly combining these chosen $K$ rows of $\bm{F}$. Thus,
\begin{align}
K &\geq 1 + Max\ No.\ of \ 0s \ in \ any\ column \ of \ \bm{F} \label{eq:col-bound1}\\
& \geq 1+ Avg. \ No.\ of \ 0s \ over \ columns \ of \ \bm{F} \label{eq:col-bound2}
\end{align} 
Here the last line follows since maximum value is always greater than average. Note that if $\bar{s}$ is the average sparsity over the rows of $\bm{F}_{P \times N}$, then the average number of zeros over the columns of $\bm{F}_{P \times N}$ can be written as $\frac{(N-\bar{s})P}{N}$. Thus, from \eqref{eq:col-bound2},
\begin{equation}
 K \ \geq 1+ \frac{(N-\bar{s})P}{N}.
\end{equation}
A slight re-arrangement establishes the lower bound in Theorem \ref{thm:fundamental1}. \hfill $\blacksquare$\\ Recall that, Short-Dot achieves a column sparsity of at most $(P-K+M)$ while a hard lower bound is $(P-K+1)$ from this proof. The bound is tight for $M=1$. The bound on average row-sparsity $s \geq \frac{N}{P}\big(P-K + 1\big) $ is also tight only for $M=1$ (implicitly assuming $P$ divides $N$, since $P \ll N$). Now we tighten this bound further for $M>1$.


\subsection{Tighter Fundamental Bounds}
\begin{theorem}
\label{thm:fundamental2}
Let $M>1$. Then there exists a matrix $\bm{A}_{M \times N}$, such that any $\bm{F}_{P \times N}$ satisfying the property that any $K$ rows of $\bm{F}_{P \times N}$ can span all the rows of $\bm{A}_{M \times N}$, must also satisfy the following property:

The average sparsity over the rows of $\bm{F}_{P \times N}$ is lower bounded as
\begin{equation}
\bar{s} > \frac{N}{P}\big(P-K +M\big) - \frac{M^2}{P}\binom{P}{K-M+1}
\label{eq:fundamental-nonasymptotic}
\end{equation}
Moreover, if $N$ is sufficiently large, such that $M^2 \binom{P}{K-M+1} =o(N)$, then  the average sparsity over the rows of $\bm{F}_{P \times N}$ is lower bounded as
\begin{equation}
\bar{s} > \frac{N}{P}(P-K +M) - o\Big(\frac{N}{P}\Big)
\label{eq:fundamental-asymptotic}
\end{equation}
\end{theorem}
Note that the second term in the lower bound in \eqref{eq:fundamental-nonasymptotic} does not depend on $N$. Thus, if $N$ is sufficiently larger than $P$ and $M$, the second term in the lower bound becomes negligible compared to the first term, and the first term is precisely what Short-Dot can achieve. Thus, from this lower bound, we can conclude that when $N$ is large, Short-Dot is near optimal. 

Before proceeding with the proof, we give a basic intuition on the proof technique. We basically divide the columns of $\bm{F}_{P \times N}$ into two groups, one with at most $(K-M)$ zeros, and other with more than $(K-M)$ zeros. Then we show that there exist matrices $\bm{A}_{M \times N}$ such that the number of columns in the latter group, \textit{i.e.},  with more than $(K-M)$ zeros is bounded, and this in turn bounds the average sparsity. Now we formally prove the theorem.

\textit{Proof:} Let us denote the number of columns of $\bm{F}_{P \times N}$ with more than $(K-M)$ zeros as $\lambda$. We will show later in \textit{Lemma 3} that $\lambda < M\binom{P}{K-M+1}$. Now, compute the average number of zeros over the columns of $\bm{F}$. The columns of $\bm{F}$ can be divided into two groups : $\lambda$ columns with greater than $(K-M)$ zeros and $(N-\lambda)$ columns with at-most $(K-M)$ zeros.  Recall from \eqref{eq:col-bound1}, that if $\bm{A}$ is chosen such that every column has at-least one non-zero entry, then the maximum number of zeros in any column of $\bm{F}$ is upper bounded by $(K-1)$.
Thus, the group of $\lambda$ columns can have at-most $K-1$ zeros each. Thus,
\begin{align}
\label{column_bound}
&\text{Avg. no. of 0s per column}  \leq  \frac{ (K-1)\lambda + (K-M)(N-\lambda) }{N} \nonumber \\
& = K-M + \frac{\lambda(M-1)}{N} 
 \overset{Lemma \ 3}{<} K-M + \frac{M^2 \binom{P}{K-M+1}}{N} 
\end{align} 
If $\bar{s}$ is the average sparsity of each row of $\bm{F}$, then the average zeros of each column of $\bm{F}$ is given by $\frac{(N-\bar{s})P}{N}$. Thus,
\begin{equation}
\frac{(N-\bar{s})P}{N}< (K-M)+ \frac{M^2}{N} \binom{P}{K-M+1} 
\end{equation}

After slight re-arrangement, the average sparsity of each row of $\bm{F}$ can be bounded as:
\begin{equation}
\bar{s} > \frac{N}{P}\left(P-K +M\right ) - \frac{M^2}{P} \binom{P}{K-M+1}
\end{equation}
Thus, the first part of the theorem, \textit{i.e.}, \eqref{eq:fundamental-nonasymptotic} is proved. Using the condition that $M^2\binom{P}{K-M+1}=o(N)$ in \eqref{eq:fundamental-nonasymptotic},  we can also obtain \eqref{eq:fundamental-asymptotic}. Thus,
\begin{equation}
s > \frac{N}{P}\left(P-K +M\right ) - o\left(\frac{N}{P}\right)
\end{equation} 
Thus, the theorem is proved. \hfill $\blacksquare$

Now it only remains to prove \textit{Lemma 3}.

\textit{Lemma 3:} Let $M>1$. Then there exists a matrix $\bm{A}_{M \times N}$, such that any $\bm{F}_{P \times N}$ satisfying the property that any $K$ rows of $\bm{F}_{P \times N}$ can span all the rows of $\bm{A}_{M \times N}$, must also satisfy the following property:
The number of columns $(\lambda)$ with more than $K-M$ zeros is upper bounded as $\lambda < M\binom{P}{K-M+1}$.

\textit{Proof:} Assume, $\lambda \geq M\binom{P}{K-M+1}$.  Now, a column with more than $(K-M)$ zeros will have at least $(K-M+1)$ zeros. There can be at most $\binom{P}{K-M+1}$ different patterns in which $(K-M+1)$ zeros can occur in a column of length $P$. Every column with more than $(K-M+1)$ zeros also has one of these $\binom{P}{K-M+1}$ column sparsity pattern, just with more zeros. From a pigeon-hole argument, at least one of these sparsity patterns of $(K-M+1)$ zeros will surely occur in $ \frac{\lambda}{\binom{P}{K-M+1}}$ columns or more. Let us consider the sub-matrix of $\bm{F}$, of size $P \times \frac{\lambda}{\binom{P}{K-M+1}}$, consisting of only the columns of $\bm{F}$ having $(K-M+1)$ zeros in the same locations, \textit{i.e.}, with similar sparsity pattern. Any $K$ rows of this sub-matrix of $\bm{F}$ should generate all the rows of a corresponding $M \times \frac{\lambda}{\binom{P}{K-M+1}}$ sub-matrix of the given $\bm{A}$, consisting of the same columns of $\bm{A}$ as picked in this sub-matrix of $\bm{F}$. 

There always exists a fully dense matrix $\bm{A}$ such any $M \times \frac{\lambda}{\binom{P}{K-M+1}}$ sub-matrix of $\bm{A}$ is full-rank, since $\bm{A}$ can be arbitrary. This sub-matrix of $\bm{A}$ is of rank $\min\{M , \frac{\lambda}{\binom{P}{K-M+1}} \} = M $(from assumption). Any $K$ rows of the sub-matrix of $\bm{F}$, should generate $M$ linearly independent rows of this sub-matrix of $\bm{A}$. But since the sub-matrix of $\bm{F}$ has $(K-M+1)$ rows consisting of all zeros, there is a choice of $K$ rows, such that all these zero rows are chosen, and we are only left with at most $M-1$ non-zero rows to generate $M$ linearly independent rows of $\bm{A}$. This is a contradiction. Thus, we must have $\lambda > M\binom{P}{K-M+1}$. \hfill $\blacksquare$

\section{Analysis of expected computation time for exponential tail models}
\label{sec:analysis}

\begin{figure}[ht]
\centering
\fbox{\includegraphics[height=4cm]{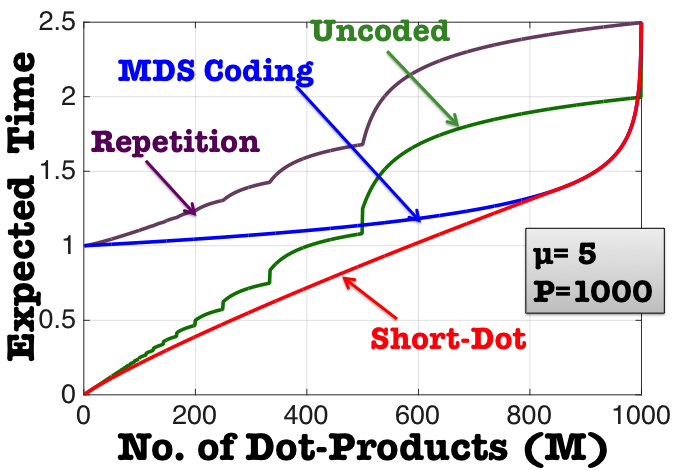}} 
\caption{Expected computation time: Short-Dot is faster than MDS when $M \ll P$ and Uncoded when $M \approx P$, and is universally faster over the entire range of $M$. For the choice of straggling parameter, Repetition is slowest. When $M$ does not exactly divide $P$, the distribution of computation time for repetition and uncoded strategies is the maximum of non-identical but independent random variables, which produce the ripples in these curves (see Appendix for details).}\label{fig0}
\end{figure}

\newcommand{\E}{{\rm I\kern-.3em E}}
We now provide a probabilistic analysis of the computation time required by Short-Dot and compare it with uncoded parallel processing, repetition and MDS coding based linear computation scheme as shown in Fig.~\ref{fig0}. We follow the shifted-exponential computation time model as described in \cite{kananspeeding}. Although the shifted exponential distribution may only be a crude approximation of the delay of real systems, we use the shifted exponential model since it is analytically tractable and allows for a fair comparison with the strategy proposed in \cite{kananspeeding}. We assume that the time required by a processor to compute a single dot-product of length $N$ be distributed as:
\begin{equation}\label{eq:dist}
\Pr(T_N \leq t)  =
\begin{cases}
      1 - \exp \left(-\mu \ \left(\frac{t}{N}-1\right) \right) , & \forall \  t \geq N  \\
      0, & \text{otherwise}
    \end{cases}
\end{equation}
Here, $\mu (> 0)$ is the ``straggling parameter'' that determines the unpredictable latency in computation time. Intuitively, the shifted exponential model states that for a task of size $N$, there is a minimum time offset proportional to $N$ such that the probability of completion of the task before that time is $0$. The probability of task completion is maximum at the time-offset and then decays with an exponential tail after that. This nature of the model might be attributed to the fact that while a processor is most likely to finish its task of size $N$ at a time proportional to $N$, but an unpredictable latency due to queuing and various other factors causes an exponential tail. For an $s$ length dot product, we simply replace $N$ by $s$ in~\eqref{eq:dist}, as suggested in \cite{kananspeeding}. The analysis of expected computation time requires closed form expressions of the $K$-th statistic which is simplistic for exponential tails. However a more thorough empirical study is necessary to establish any chosen model for straggling in a particular environment.

The expected computation time for Short-Dot is the expected value of the $K$-th order statistic of these $P$ \textit{iid} exponential random variables, which is given by:
\begin{equation}\label{eq:expected}
\E[T_{SD}] =  s \left( 1 + \frac{\log(\frac{P}{P-K})}{\mu}   \right) = \frac{(P-K+M)N}{P} \left( 1 + \frac{\log(\frac{P}{P-K})}{\mu} \right).
\end{equation}
Here,~\eqref{eq:expected} uses the fact that the expected value of the $K$-th statistic of $P$ \textit{iid} exponential random variables with parameter $1$ is  $\sum_{i=1}^P \frac{1}{i} - \sum_{i=1}^{P-K} \frac{1}{i} \approx \log(P)-\log(P-K)$ \cite{kananspeeding}. The expected computation time in the RHS of~\eqref{eq:expected} is minimized when $P-K = \Theta(M)$. This minimal expected time is $\mathcal{O}(\frac{MN}{P})$ for $M$ linear in $P$ and is $\mathcal{O}\left(\frac{MN\log(P/M)}{P }\right)$ for $M$ sub-linear in $P$.

A detailed analysis of the expected computation time for the competing strategies, \textit{i.e.}, uncoded strategy, repetition and MDS coding strategy is provided in the Appendix. Table~\ref{time-table} shows the order-sense expected computation time in the regimes where $M$ is linear and sub-linear in $P$.
\begin{table}[ht]
  \caption{Probabilistic Computation Times}
  \label{time-table}
  \centering
  \begin{threeparttable}
  \begin{tabular}{p{35mm}lll}
    \toprule
    Strategy & Expected Time &  $M$ linear in $P$ & $M$ sub-linear in $P$ \\
    \midrule
    Only one Processor & $MN\left( 1 +\frac{1}{\mu}\right)$  & $\Theta \left(MN \right)$ &  $\Theta \left(MN \right)$   \\
   Uncoded (M divides P)\tnote{2}   & $\frac{MN}{P} \left( 1 + \frac{\log(P)}{\mu}   \right) $   & $\Theta \left(\frac{MN}{P}\log(P)\right) $ & $\Theta \left( \frac{MN}{P}\log(P)\right) $ \\
    Repetition (M divides P) \tnote{2}& $N\left( 1 + \frac{M\log(M)}{P\mu}   \right) $ & $\Theta \left(\frac{MN}{P} \log(P)\right) $ & $\Theta \left(N\right)$ \\
   MDS    & $N\Big( 1 + \frac{\log \big(\frac{P}{P-M}\big)}{\mu} \Big)$  & $\Theta(N)$ & $\Theta(N)$     \\
   Short-Dot & $ \frac{N(P-K+M)}{P} \Big( 1 + \frac{\log \left(\frac{P}{P-K}\right)}{\mu} \Big)  $ & $\mathcal{O}(\frac{MN}{P}) $ & $ \mathcal{O}\left( \frac{MN}{P}\log \left(\frac{P}{M}\right)  \right)   $\\
    \bottomrule
  \end{tabular}
  \begin{tablenotes}
            \item[2] Refer to Appendix for more accurate analysis taking integer effects into account
        \end{tablenotes}
   \end{threeparttable}
\end{table}

Note that in the regime where $M$ is linear in $P$, Short-Dot outperforms Uncoded Strategy by a factor diverging to infinity for large $P$. Similarly, in the regime where $M$ is sub-linear in $P$, Short-Dot outperforms MDS coding strategy by a factor that diverges to infinity for large $P$. Thus Short-Dot universally outperforms all its competing strategies over the entire range of $M$. 

Now we explicitly provide a regime, where the speed-ups from Short-Dot diverges to infinity for large $P$, in comparison to all three competing strategies - MDS Coding, Repetition or Uncoded strategies.
\begin{theorem}
Suppose $M$ scales as $\frac{P}{\log{P}}$. Then, Short-Dot with $K=P-\frac{M}{2}$ has an expected computation time (scaled by $N$) as  $\frac{\E[T_{SD}]}{N}=O(\frac{\log(\log{P})}{\log{P}})$ that decays to $0$ as $P \to \infty$. In contrast, the expected computation time (scaled by $N$) for MDS coding, repetition and uncoded strategies scale as $\Omega(1)$ and thus do not decay to $0$ as $P \to \infty$. 
\end{theorem}

\textit{Proof:} For the proof of this theorem, we simply substitute the values of $M$ and $K$ in the expressions of expected computation time as follows. We let $M=\frac{P}{\log P}$ for all the strategies. 
For uncoded strategy, we thus obtain,
\begin{equation}
\frac{\E[T_{UC}]}{N} = \frac{M}{P} \left( 1 + \frac{\log(P)}{\mu}   \right) = \frac{1}{\log{P}}\left( 1 + \frac{\log(P)}{\mu}   \right) \geq \frac{1}{\mu} = \Omega (1)
\end{equation}
For repetition, we obtain,
\begin{equation}
\frac{\E[T_{Rep}]}{N} = \Big( 1 + \frac{M\log(M)}{P\mu}   \Big)  \geq 1 = \Omega (1)
\end{equation}
For MDS Coding based linear computation, we obtain,
\begin{equation}
\frac{\E[T_{MDS}]}{N} = 1 + \frac{\log\left(\frac{P}{P-M}\right)}{\mu} = 1 + \frac{\log\left(\frac{\log{P}}{\log{P}-1}\right)}{\mu}  \geq 1 = \Omega (1)
\end{equation}
\begin{figure}[ht]
\centering
\fbox{\includegraphics[height=3.1cm]{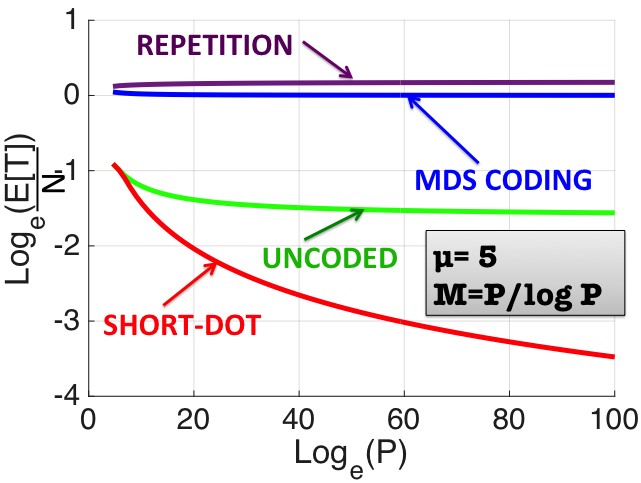}}
\hspace{0.5cm}
\fbox{\includegraphics[height=3.1cm]{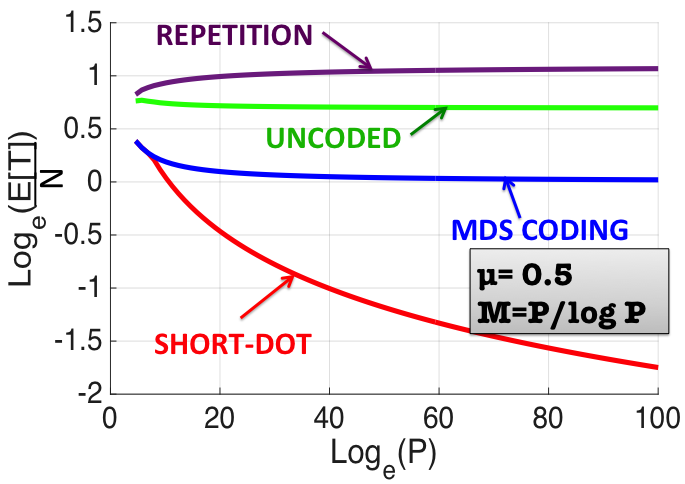}}
\caption{Log of Expected Computation Time scaled by $N$ with $\log(P)$ where $P$ is the number of processors, under $M=P/\log{P}$: Short-Dot offers speed-ups compared to uncoded, repetition and MDS coding that diverge for large $P$.  }\label{fig:comp_time_scaling}
\end{figure} 
Now, we consider the Short-Dot strategy with $K=P-\frac{M}{2} = P- \frac{P}{2\log{P}}$. Note that the inequality $K > M$ is satisfied for $\log{P}>\frac{3}{2}$. Now let us calculate the expected computation time for Short-Dot.
\begin{equation}
\frac{\E[T_{SD}]}{N} = \frac{(P-K+M)}{P} \Big( 1 + \frac{\log \big(\frac{P}{P-K}\big)}{\mu} \Big)  = \frac{3}{2\log{P}}\left( 1 + \frac{\log(2\log{P})}{\mu}   \right) \leq \mathcal{O}\big( \frac{\log(\log P)}{\log{P}}\big)
\end{equation}
Thus, the speed-up offered by Short-Dot in this regime is $\frac{\log{P}}{\log(\log P)}$, and thus diverges to infinity for large $P$, as illustrated in Fig.~\ref{fig:comp_time_scaling}.

\section{Encoding and Decoding Complexity} 

\subsection{Encoding Complexity:} Even though encoding is a pre-processing step (since $\bm{A}$ is assumed to be given in advance), we include a complexity analysis for the sake of completeness. Recall from Section \ref{sec:shortdot} that we first choose an appropriate matrix $\bm{B}$ of dimension $P \times K$, such that every $K \times K$ square sub-matrix is invertible and all $(K-M)\times(K-M)$ sub-matrices in the last $(K-M)$ columns are invertible. Now, for each of the $N$ columns of the given matrix $\bm{A}$, we perform the following.
\begin{algorithmic}
\State  \textbf{Set } $ \ \ \  \ U \gets (\{ (j-1), \ldots, (j+K-M-1) \} \mod P) +1 $\\
 \hfill $\rhd$ The set of $(K-M)$ indices that are 0 for the $j$-th column of $\bm{F}$
\State  \textbf{Set} $\ \ \ \bm{B}^U \gets \text{Rows of } \bm{B} \text{ indexed by $U$}$
\State  \textbf{Solve for} $ \bm{z}: \ \ \ \ \  (\bm{B}^U_{cols \ M+1:K})[\ \bm{z}\ ]=  -  \ \bm{B}^U_{cols \ 1:M}[\bm{A}_j]  $  \hfill $\rhd$ $\bm{z}_{(K-M) \times 1}$ is a row vector.
\State   \textbf{Set} $ \ \ \ \ \ \bm{F}_{j} = \bm{B}[\bm{A}_j^T \ | \bm{z}^T \ ]^T $ \hfill $\rhd$ $\bm{F}_{j}$ is a column vector ( $j$-th col of $\bm{F}$)
\end{algorithmic}

For each of the $N$ columns, the encoding requires a matrix inversion of size $(K-M) \times (K-M)$ to solve a linear system of equations, a matrix-vector product of size $(K-M)\times M $ and another matrix vector product of size $ P \times K$. \\
The naive encoding complexity is therefore $\mathcal{O}(N((K-M)^3 + (K-M)M + PK))$. Note that effectively there are only $N/P$ different column sparsity patterns for this particular design discussed in this paper. Thus, there are effectively $N/P$ unique $\bm{B}^U$s , and thus $N/P$ unique matrix inversions can suffice for all the $N$ columns, as sparsity pattern is repeated. Thus, the complexity can be reduced to
$\mathcal{O}(\frac{N}{P}(K-M)^3 + (K-M)MN + PKN)= \mathcal{O}(\frac{N}{P}(K-M)^3 +2PKN )$

This is higher than MDS coding based linear computation that has an encoding complexity of $\mathcal{O}(NMP))$, but it is only a one-time cost that provides savings in online steps (as discussed earlier in this section). 
\subsubsection{Reduced Complexity using Vandermonde matrices:} The encoding complexity can be reduced further for special choices of the matrix $\bm{B}$. Let us choose $\bm{B}$ to be a Vandermonde matrix as given by 
\begin{equation}
\label{eq:vandermonde}
\bm{B}=\begin{bmatrix} h_1^{K-1}  & \dots & h_1 & 1\\ 
h_2^{K-1}  & \dots & h_2 & 1\\ 
\vdots & \ddots & \vdots & \vdots\\
h_P^{K-1}  & \dots & h_P & 1
\end{bmatrix}
\end{equation}
Here, $ h_1, h_2, \dots, h_K \in \mathbb{R} $, and are all distinct. This matrix $\bm{B}$ satisfies all the requirements of the encoding matrix. All $K \times K$ sub-matrices of $\bm{B}$ are invertible, and all $(K-M) \times (K-M)$ sub-matrices in the last $(K-M)$ columns are also invertible. Thus, this matrix can be used to encode the matrix $\bm{F}$. For each of the $N$ columns of $\bm{F}$, the encoding requires solving a linear system of equations for $\bm{z}$, as given by:
\begin{equation}
(\bm{B}^U_{cols \ M+1:K})[\ \bm{z}\ ]=  -  \ \bm{B}^U_{cols \ 1:M}[\bm{A}_j] 
\end{equation} 
Here $U$ denotes a set of $(K-M)$ indices $\in \{1,2,\dots,P\}$.\\
The matrix-vector product $\bm{B}^U_{cols \ 1:M}[\bm{A}_j]$ is equivalent to the evaluation of a polynomial of degree $(K-1)$ with the $K$ co-efficients as $[\bm{A}_j^T  \bm{0}_{(K-M)\times 1}  ]$ at $(K-M)$ arbitrary points given by $\{h_l| l \in U\}$. Once this product is obtained, the linear system of equations reduces to the interpolation of the $(K-M)$ unknown co-efficients of a polynomial of degree $(K-M-1)$ (which is $\bm{z}$), from its value at $(K-M)$ arbitrary points as given by $\{h_l| l \in U\}$. Once $\bm{z}$ is obtained, we perform the following operation.
\begin{equation}
\bm{F}_{j} = \bm{B}[\bm{A}_j^T \ | \bm{z}^T \ ]^T
\end{equation}
This step is equivalent to the evaluation of a polynomial of degree $(K-1)$ at $P$ points given by $\{h_l| l=1,2,\dots P \}$. Thus we decompose our encoding problem for each column of $\bm{A}$ into a bunch of polynomial evaluation and interpolation problems, all of degree less than $P$. Now, from \cite{kung1973fast}, \cite{li2000arithmetic}, we know that both the interpolation and the evaluation of a polynomial of degree less than $P$, at $P$ arbitrary points is $\mathcal{O}(P \log^2(P))$.
 Thus, the complexity of encoding is $\mathcal{O}(NP \log^2(P))$.
\subsection{Decoding Complexity:}
During decoding, we get $K$ dot-products from the first $K$ processors out of $P$. We then perform the following operations.
\begin{algorithmic}
\State  \textbf{Set }  $\ \ \  \ V \gets \text{Indices of the }  \text{ processors that finished first}$
\State  \textbf{Set} $\ \ \ \ \bm{B}^V \gets \text{Rows of } \bm{B} \text{ indexed by $V$}$
\State  \textbf{Set} $\ \ \ \ \bm{v}_{K \times 1} \gets [\langle \bm{f}_i^{S_i} , \bm{x}^{S_i} \rangle , \ \forall \ i \ \in V]$ \hfill $\rhd$ Col Vector of outputs  from first $ K $ processors
\State \textbf{Solve for } $\bm{w}_{K \times 1}: \ \ \ \  [(\bm{B}^V)]\bm{w}= \bm{v}$
\State \textbf{Output:} $\bm{Ax}=[w_1, w_2, \dots w_M]^T $ \hfill $\rhd$ First $M$ values of $\bm{w}$
\end{algorithmic}

We solve a system of $K$ linear equations in $K$ variables and use only $M$ values of the obtained solution vector.  Thus, effectively we do a single matrix inversion of size $K \times K$ followed by a matrix-vector product of size $K \times M$. The decoding complexity of Short-Dot is thus $\mathcal{O}(K^3 + KM)$ which does not depend on $N$ when $M,K \ll N$. This is nearly the same as  $\mathcal{O}(M^3 + M^2)$ complexity of MDS coding based linear computation. 

\subsubsection{Reduced Complexity using Vandermonde matrices:}
Similar to encoding, using Vandermonde matrices can reduce the decoding complexity further. As already discussed, we choose the encoding matrix $\bm{B}$ as a Vandermonde matrix as described in \eqref{eq:vandermonde}. The decoding problem consists of solving a  a system of $K$ linear equations in $K$ variables.
\begin{equation}
[(\bm{B}^V)]\bm{w}= \bm{v}
\end{equation}
Here $V$ is a set of $K$ indices $\in \{1,2,\dots,P\}$. The problem of finding $\bm{w}$ is equivalent to the interpolation of the co-efficients of a polynomial of degree $(K-1)$, from its values at $K$ arbitrary points given by $\{h_l|l \in V\}$. Again, from \cite{kung1973fast}, \cite{li2000arithmetic}, the interpolation of a polynomial of degree $(K-1)$, at $K$ arbitrary points can be done in $\mathcal{O}(K \log^2(K))$, which thus becomes the decoding complexity.

\section{Experimental Results}\label{sec:experiments}
We perform experiments on computing clusters at CMU to test the computational time. We use HTCondor~\cite{HTCondor} to schedule jobs simultaneously among the $P$ processors. We compare the time required to classify $10000$ handwritten digits of the MNIST~\cite{lecun1998mnist} database, assuming we are given a trained $1$-layer Neural Network. We separately trained the Neural network using training samples, to form a matrix of weights, denoted by $\bm{A}_{10 \times 785}$. For testing, the multiplication of this given $10 \times 785$ matrix, with the test data matrix $\bm{X}_{785 \times 10000}$ is considered. The total number of processors was $20$. 

\begin{figure}[ht]
\centering
\fbox{\includegraphics[ height=3.1cm]{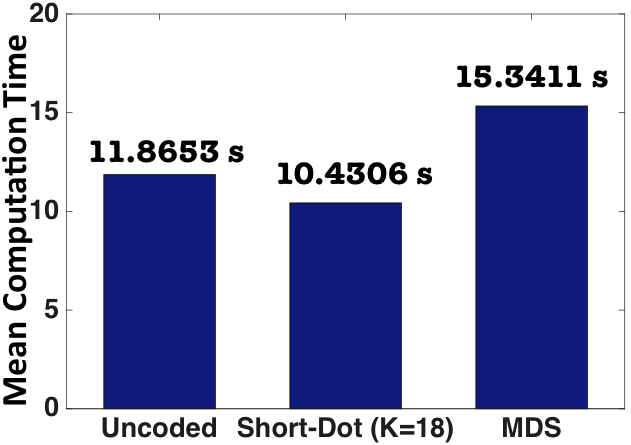}}
\hspace{0.2cm}
\fbox{\includegraphics[ height=3.1cm]{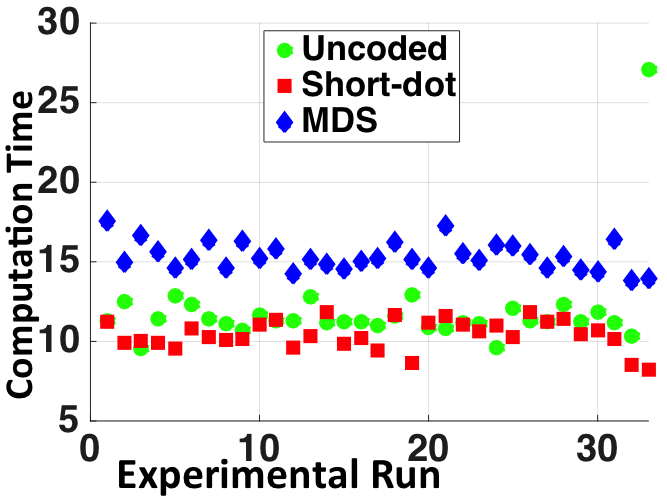}}
\caption{Experimental results: (Left) Mean computation time for Uncoded Strategy, Short-Dot (K=18) and MDS codes: Short-Dot is faster than MDS by $32\%$ and Uncoded by $12\%$. (Right) Scatter plot of computation time for different experimental runs: Short-Dot is faster most of the time. }\label{sim_fig}
\end{figure}

Assuming that $\bm{A}_{10 \times 785}$ is encoded into $\bm{F}_{20 \times 785}$ in a pre-processing step, we store the rows of $\bm{F}$ in each processor apriori. Now portions of the data matrix $\bm{X}$ of size $s \times 10000$ are sent to each of the $P$ parallel processors as input. We also send a C-program to compute dot-products of length $s=\frac{N}{P}(P-K+M)$ with appropriate rows of $\bm{F}$ using command \textit{condor-submit}. Each processor outputs the value of one dot-product. The computation time reported in Fig. \ref{sim_fig} includes the total time required to communicate inputs to each processor, compute the dot-products in parallel, fetch the required outputs, decode and classify all the $10000$ test-images, based on $35$ experimental runs.
\begin{table}[t]
  \caption{Experimental computation time of $10000$ dot products ($N=785 , M=10 , P=20 $)}
 \centering
 \begin{tabular}{p{15mm}p{18mm}p{10mm}p{15mm}p{22mm}p{23mm}}
    \toprule
  Strategy  & Parameter $K$ & Mean & STDEV & Minimum Time & Maximum Time\\
    \midrule
  Uncoded & 20 & 11.8653 & 2.8427 & 9.5192  & 27.0818 \\
  \midrule
  Short-Dot & 18  & 10.4306 & 0.9253 & 8.2145 & 11.8340\\
  \midrule
 MDS & 10  & 15.3411 & 0.8987 & 13.8232 & 17.5416 \\
 \hline
  \end{tabular}\label{results_2}
\end{table}

\textbf{Key Observations:} (See Table~\ref{results_2} for detailed results). Computation time varies based on nature of straggling, at the particular instant of the experimental run. Short-Dot outperforms both MDS and Uncoded, in mean computation time. Uncoded is faster than MDS since per-processor computation time for MDS is larger, and it increases the straggling, even though MDS waits for only for $10$ out of $20$ processors. However, note that Uncoded has more variability than both MDS and Short-Dot, and its maximum time observed during the experiment is much greater than both MDS and Short-Dot. The classification accuracy was $85.98 \%$ on test data. 

{\color{black}{\paragraph{Comment:} The experimental times are quite high due to some limitations of the experimental platform used. The time includes some overhead to start the cluster, and communicate data in the form of text files to all the processors, and also collect the output data files back from all the processors. The read time also depends on the size of file to be read. Currently, we are looking at performing these experiments in alternate distributed computing platforms, with better communication protocols.}}

\section{Discussion}
\label{sec:discussion}
\subsection{Storage and Communication benefits of Shorter Dot Products:} { \color{black}{The major advantage of using Short-Dot codes over the MDS coding strategy in \cite{kananspeeding} is that the length of the pre-stored vectors (rows of $\bm{F}$) as well as the communicated input (portions of $\bm{x}$) is shorter than $N$. It is thus applicable when processing units have limitations of memory, and it is not possible to pre-store the long vectors of length $N$. Short-dot also has advantages over \cite{kananspeeding} in systems where the principle bottlenecks in computation time is in communicating the input $\bm{x}$ to all the processors, and it may not be feasible to broadcast (multicast) $\bm{x}$ to all processors at the same time. Thus, it is also useful in applications where communication costs are predominant over computation costs.} }

\subsection{Errors instead of erasures:}
While we focus on the problem of erasures in this paper, Short-Dot can also be used to correct errors. Consider the scenario  when instead of straggling or failures, some processors return entirely faulty or garbage outputs, in a distributed system and we do not know which of the outputs are erroneous. We argue from coding theoretic arguments that Short-Dot codes designed to tolerate $(P-K)$ stragglers, can also correct $\lfloor \frac{(P-K)}{2}\rfloor $ errors. First observe that if the code can tolerate $(P-K)$ stragglers, then the Hamming Distance between any two code-words should at least be $(P-K+1)$. Hence, the number of errors that can be corrected is  $\lfloor \frac{(\text{Hamming Distance -1})}{2}\rfloor $ which is $\lfloor \frac{(P-K)}{2}\rfloor $. The same result can also be derived by recasting the decoding problem as a sparse reconstruction problem, and borrowing ideas from standard compressive sensing literature \cite{candes2005decoding} which also yields a concrete, decoding algorithm. The problem reduces to an $l_0$ minimization problem, which can be relaxed into an $l_1$ minimization, or solved using alternate sparse reconstruction techniques, under certain constraints on the encoding matrix $\bm{B}$.


\subsection{More dot-products than processors}
While we have presented the case of $M<P$ here, Short-Dot easily generalizes to the case where $M \geq P$. The matrix can be divided horizontally into several chunks along the row dimension (shorter matrices) and Short-Dot can be applied on each of those chunks one after another. Moreover if rows with same sparsity pattern are grouped together and stored in the same processor initially, then the communication cost is also significantly reduced during the online computations, since only some elements of the unknown vector $\bm{x}$ are sent to a particular processor.
  
\textbf{Acknowledgments:} Systems on Nanoscale Information fabriCs (SONIC), one of the six SRC STARnet Centers, sponsored by MARCO and DARPA. We also acknowledge NSF Awards 1350314, 1464336 and 1553248. S Dutta also received Prabhu and Poonam Goel Graduate Fellowship.


\bibliographystyle{unsrt}
\bibliography{sample}


\section{Appendix}
We now provide a probabilistic analysis of the computational time required by Short-Dot and compare it with uncoded parallel processing, repetition and MDS code based linear computation as shown in Fig. \ref{fig0}. We assume that the time required by a processor to compute a single dot-product follows an exponential distribution and is independent of other parallel processors. 

Let us assume, the time required to compute a single dot-product of length $N$, follow the distribution:-
\begin{equation}
\Pr(T_N \leq t) = F(t) =\begin{cases}
      1 - \exp \left(-\mu \ \left(\frac{t}{N}-1\right) \right) , & \forall \  t \geq N  \\
      0, & \text{otherwise}
    \end{cases} 
\end{equation}
Here, $\mu (>0)$ is a straggling parameter, that determines the ``unpredictable latency'' in computation time. We also assume, that if the length of the dot-product is $s$ where $s$ is the sparsity of the vector, the probability distribution of the computational time varies as:-
\begin{equation}
\Pr(T_s \leq t) =F\left(\frac{Nt}{s}\right) =\begin{cases}
      1 - \exp \left(-\mu \ \left(\frac{t}{s}-1\right) \right) , & \forall \  t \geq s   \\
      0, & \text{otherwise}
    \end{cases} 
\end{equation}
Now we derive the expected computation time using our proposed strategy and compare it with existing strategies in the regimes where the number of dot-products $M$ is linear and sub-linear in $P$.

Table \ref{time-table} shows the order-sense expected computation time in the regimes where $M$ is linear and sub-linear in $P$. 

\subsection{Proposed Strategy -- Short-Dot:}
The computation time over each of the $P$ processors behaves as \textit{iid} exponential random variables following the distribution:-
\begin{equation}
\Pr(T_s \leq t) =F\left(\frac{Nt}{s}\right) = 1 - \exp\left(-\mu \ \left(\frac{t}{s}-1\right)\right) \ \ \forall \  t \geq s \ .
\end{equation}
Now, the expected computation time is the expected value of the $K$-th order statistic of these $P$ \textit{iid} exponential random variables, which is given by:-
\begin{equation}
\E [T_{SD}] =  s \left( 1 + \frac{\log(\frac{P}{P-K})}{\mu}   \right) = \frac{(P-K+M)N}{P} \left( 1 + \frac{\log(\frac{P}{P-K})}{\mu} \right)
\end{equation}
Here we use the result (from  \cite{kananspeeding}) that the $K-$ th order statistic of $P$ exponential random variables that are \textit{iid} as $\sim \exp(-T) \  \forall \ T \ \geq 0  $  is given by $$\sum_{i=1}^P \frac{1}{i} - \sum_{i=1}^{P-K} \frac{1}{i}$$ For large $P$ and $K < P$, we can approximate the following: 
\begin{equation}
\sum_{i=1}^P \frac{1}{i} - \sum_{i=1}^{P-K} \frac{1}{i} \approx \log(P)-\log(P-K)
\end{equation}

Note that the expected computation time is minimized when $K =P- \Theta(M)$, and is given by:-
\begin{equation}
\label{short_dot_time}
\E [T_{SD}^{*}] = \mathcal{O}\left(\frac{MN}{P}\left(1 + \frac{\log(P/M)}{\mu}  \right)\right)
\end{equation}
If $M=\Theta(P)$, the expected time is $\mathcal{O}(\frac{MN}{P})$. If $M=o(P)$, the expected time is $\mathcal{O}\left(\frac{MN\log(P/M)}{P }\right)$. Note that $s=\frac{(P-K+M)N}{P}$ is actually an upper bound on the length of each dot-product achieved using Short-Dot. Thus the expression obtained in (\ref{short_dot_time}) is an upper bound for the actual expected computation time. Thus we use $\mathcal{O}(.)$ instead of $\Theta(.)$.

\subsection{Existing Strategies}
\paragraph{One Single Processor:}
For one single processor to compute all $M$ dot-products of length $N$, the computation time is distributed as 
\begin{equation}
\Pr(T_{NM} \leq t) =F\big(t/M\big) = 1 - \exp \left(-\mu \ \left(\frac{t}{NM}-1\right) \right) \ \ \forall \  t \geq NM   
\end{equation}
Thus, the expected computation time can be easily derived to be
\begin{equation}
\E [T_{1P}]=MN\left( 1 +\frac{1}{\mu}\right)
\end{equation}
\paragraph{Uncoded - Divide into $P$ parts and wait for all:}
Now, consider an uncoded strategy where the computation is simply divided into $P$ dot-products and sent to $P$ processors. We assume that each processor is sent only one dot-product at a time. We wait for all the processors to finish computation. Note that integer effects arise when $M$ does not exactly divide $P$. Some rows can be divided among $\left \lceil{\frac{P}{M}}\right \rceil $ processors, while the remaining are divided among $\left \lfloor{\frac{P}{M}}\right \rfloor $ processors. Let $m_1$ and $m_2$ denote the number of rows that get $\left \lceil{\frac{P}{M}}\right \rceil $ processors and $\left \lfloor{\frac{P}{M}}\right \rfloor $ processors respectively. Clearly the values can be obtained by solving:-
\begin{equation}
\begin{bmatrix} 
  1     & 1\\ 
  \left \lceil{\frac{P}{M}}\right \rceil  & \left \lfloor{\frac{P}{M}}\right \rfloor  
\end{bmatrix}\begin{bmatrix} 
  m_1  \\ 
 m_2  
\end{bmatrix} = \begin{bmatrix} 
  M  \\ 
 P  
\end{bmatrix}
\end{equation}

Now, we have two groups of exponential variables: one group consisting of  $m_1\left \lceil{\frac{P}{M}}\right \rceil $ \textit{iid} exponential random variables of task size $\frac{N}{\left \lceil{\frac{P}{M}}\right \rceil} $ , and another group consisting of  $m_2\left \lfloor{\frac{P}{M}}\right \rfloor $  \textit{iid} exponential random variables of task size $\frac{N}{\left \lfloor{\frac{P}{M}}\right \rfloor} $.
The two groups are independent of each other.
Note that we assume that $N$ is large compared to $P$ and is divisible by $P,\left \lfloor{\frac{P}{M}}\right \rfloor, \left \lfloor{\frac{P}{M}}\right \rfloor $, so that the integer effects with respect to $N$ do not appear and the plots can be scaled with respect to $N$ for ease of understanding. 

The expected computation time is thus given by the expectation of the maximum of all these $P=m_1\left \lceil{\frac{P}{M}}\right \rceil + m_2\left \lfloor{\frac{P}{M}}\right \rfloor $ exponential random variables.

\begin{multline}
\Pr( T_{UC} \leq t)=  \left( 1- \exp \left(-\mu  \left( \frac{\left \lceil{\frac{P}{M}}\right \rceil t}{N} -1 \right) \right)        \right)^{m_1 \left \lceil{\frac{P}{M}}\right \rceil} \times \\ \left( 1- \exp \left(-\mu \left( \frac{\left \lfloor{\frac{P}{M}}\right \rfloor  t}{N} -1 \right) \right)        \right)^{m_2 \left \lfloor{\frac{P}{M}}\right \rfloor } 
 \forall \ t \geq \frac{N}{\left \lfloor{\frac{P}{M}}\right \rfloor}
\end{multline}

The expectation is thus obtained as 
\begin{equation}
\E [T_{UC}]=\int_{0}^{\infty} \left(1- \Pr( T_{UC} \leq t) \right) dt
\end{equation}

This expression is numerically computed using MATLAB and plotted in the plot of theoretical computation time in Fig.~\ref{fig0}. When $M$ divides $P$ exactly, the expressions are simpler. The computation time for each processor is distributed as 
\begin{equation}
\Pr(T_{UC} \leq t) =F\big(t/M\big) = 1 - \exp \left(-\mu \ \left(\frac{Pt}{MN}-1\right) \right) \ \ \forall \  t \geq NM/P 
\end{equation}
The expected computation time is the maximum of $P$ such independent and identically distributed random variables, as given by:-
\begin{equation}
\E [T_{UC}]=\frac{MN}{P}\left( 1 +\frac{\log(P)}{\mu}\right)
\end{equation}
The expected time for this uncoded strategy is $\Theta\left(\frac{MN\log(P)}{P }\right)$ regardless of whether $M$ is linear or sub-linear in $P$. Our strategy Short-Dot thus offers a speed-up of $\Omega(\log(P))$ in expected computation time when $M$ is linear in $P$, as mentioned in \eqref{short_dot_time}, and thus outperforms by a factor that diverges to infinity for large $P$.

\paragraph{Repetition:}
When a $(P,M)$ repetition strategy is used, we separate the matrix into $M$ rows and repeat each row $P/M$ times, so as to obtain a total of $P$ tasks. Note that integer effects arise when $M$ does not exactly divide $P$. Some rows are repeated $\left \lceil{\frac{P}{M}}\right \rceil $ times, while the remaining are repeated $\left \lfloor{\frac{P}{M}}\right \rfloor $ times. Let $m_1$ and $m_2$ denote the number of rows that are repeated $\left \lceil{\frac{P}{M}}\right \rceil $ times and $\left \lfloor{\frac{P}{M}}\right \rfloor $ times respectively. Clearly the values can be obtained by solving:-
\begin{equation}
\begin{bmatrix} 
  1     & 1\\ 
  \left \lceil{\frac{P}{M}}\right \rceil  & \left \lfloor{\frac{P}{M}}\right \rfloor  
\end{bmatrix}\begin{bmatrix} 
  m_1  \\ 
 m_2  
\end{bmatrix} = \begin{bmatrix} 
  M  \\ 
 P  
\end{bmatrix}
\end{equation}

Now, the minimum of $\left \lceil{\frac{P}{M}}\right \rceil $ (or similarly $\left \lfloor{\frac{P}{M}}\right \rfloor $ )  \textit{iid} exponential random variables is also exponential with parameter scaled by $\left \lceil{\frac{P}{M}}\right \rceil $ (or similarly $\left \lfloor{\frac{P}{M}}\right \rfloor $ ). The expected computation time is thus given by the expectation of the maximum of $m_1$ independent exponential variables with parameter scaled by $\left \lceil{\frac{P}{M}}\right \rceil $ and $m_2$ independent exponential variables with parameter scaled by $\left \lfloor{\frac{P}{M}}\right \rfloor $. 

\begin{multline}
\Pr( T_{Rep} \leq t)= \left( 1- \exp \left(-\mu \left \lceil{\frac{P}{M}}\right \rceil \left( \frac{t}{N} -1 \right) \right)        \right)^{m_1} \times \\ \left( 1- \exp \left(-\mu \left \lfloor{\frac{P}{M}}\right \rfloor \left( \frac{t}{N} -1 \right) \right)        \right)^{m_2} \ \forall \ t \geq N
\end{multline}

\begin{figure}[ht]
\centering
\fbox{\includegraphics[width=11cm]{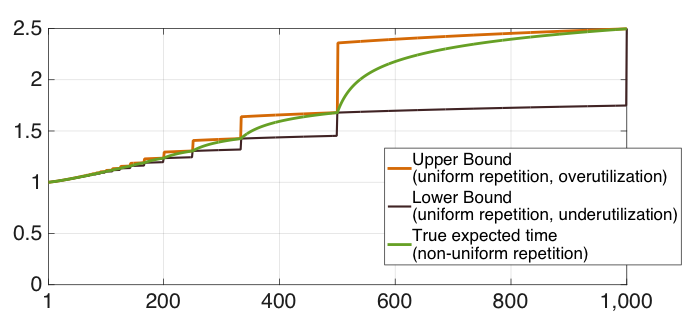}}
\caption{Theoretical Plot of expected computation time of repetition taking integer effects into account:  straggling parameter $\mu=5$, total processors $P=1000$  and number of dot-products $M$ is varied from $1$ to $P$.}\label{repetition_fig}
\end{figure}

The expectation is thus obtained as 
\begin{equation}
\E [T_{Rep}]=\int_{0}^{\infty} \left(1- \Pr( T_{Rep} \leq t) \right) dt
\end{equation}

This expression is computed using MATLAB in the plot of theoretical expected computation time (Fig.~\ref{fig0}). 
When $M$ exactly divides $P$, the analysis is simpler, and the two types of exponential distributions are identical. 
Following an analysis similar to [1], it simplifies to the expectation of the maximum of $M$ \textit{iid} exponential random variables, each of which is the minimum of $P/M$ \textit{iid} exponential random variables.
\begin{equation}
\E [T_{Rep}]=N\left( 1 + \frac{M\log(M)}{P\mu}   \right)
\end{equation}
When $M$ is linear in $P$, the expected computation time is $\Theta(\frac{MN}{P} \log (P))$ while our strategy achieves $\mathcal{O}(N)$ in this regime. When $M$ is sub-linear in $P$, the expected computation time is $\Theta(N)$ while Short-Dot achieves $\mathcal{O}\left(\frac{MN\log(P/M)}{P }\right)$ that offers speed-up by a factor diverging to infinity.
\paragraph{MDS codes-based strategy:}
The matrix is separated into $M$ rows and coded into $P$ rows using a $(P,M)$ MDS code. Thus, each processor effectively computes a dot-product of length $N$. We have to wait for any $M$ processors to finish. Assuming the computation of each processor is independent, following an analysis similar to [1], we obtain that,
\begin{equation}
\E [T_{MDS}]=N\left( 1 + \frac{\log(P)}{\mu} - \frac{\log(P-M)}{\mu}  \right)
\end{equation}
When $M$ is linear in $P$, the expected computation time is $\Theta(N)$ as compared to our strategy that achieves $\mathcal{O}(MN/P) $.  However, in the regime where $M$ is sub-linear in $P$, the expected computation time is also $\Theta(N)$ while our strategy achieves $\mathcal{O}\left(\frac{MN\log(P/M)}{P}\right)$, and thus outperforms MDS codes by a factor that diverges to infinity for large $P$.





\end{document}